# Default Risk Modeling Beyond the First-Passage Approximation: Extended Black-Cox Model.

Yuri A. Katz[a] and Nikolai V. Shokhirev[b]

P.O. Box 66, Norwalk, CT 06853

We develop a generalization of the Black-Cox structural model of default risk. The extended model captures uncertainty related to firm's ability to avoid default even if company's liabilities momentarily exceeding its assets. Diffusion in a linear potential with the *radiation* boundary condition is used to mimic a company's default process. The exact solution of the corresponding Fokker-Planck equation allows for derivation of analytical expressions for the cumulative probability of default and the relevant hazard rate. Obtained closed formulas fit well the historical data on global corporate defaults and demonstrate the split behavior of credit spreads for bonds of companies in different categories of speculative-grade ratings with varying time to maturity. Introduction of the finite rate of default at the boundary improves valuation of credit risk for short time horizons, which is the key advantage of the proposed model. We also consider the influence of uncertainty in the initial distance to the default barrier on the outcome of the model and demonstrate that this additional source of incomplete information may be responsible for non-zero credit spreads for bonds with very short time to maturity.

PACS numbers: 89.65Gh, 05.40.Jc, 05.90.+m

---

[a] yk.qubit@gmail.com

[b] nikolai.shokhirev@gmail.com

**I. Introduction**

Demand for better economic models as well as the deep analogy between phenomena in physics of complex system and economics draws attention of a quickly growing number of physicists [1]. Most of the research in econophysics concerns the financial markets. In fact, the first model of the stochastic time evolution of the system is due to Bachelier [2] who developed the theory of essentially Brownian motion to evaluate European stock options (five years before the 1905 paper of Einstein [3]). The price of the option is derived from the present price of the stock and depends on the market risk due to the uncertainty in a price change over time. In the Bachelier's model the time evolution of the stock price follows the Wiener process, which proves to be an efficient approximation for irregular time behavior of numerous variables in physics and economics.

One should distinguish market risk from credit risk or risk of default. The company defaults if it fails to pay due obligations. Therefore, credit risk is largely due to the uncertainty in a time evolution of the firm's aggregate asset value $V$. Behavior of many other determinants of the probability of default (PoD), e.g., the value of firm's short- and long-term liabilities $L$, the expected rate of assets growth $\mu$, the risk-free interest yield $r$, and the volatility of its assets $\sigma$ may be also unknown and complicate the analysis. Note that $\sigma\sqrt{\Delta t}$ is the standard deviation of the percentage change of $V$ in a short time interval $\Delta t$. Starting with Merton's "structural" model, the time-behavior of $V$ is modeled as the geometric Brownian motion [4]. Additionally, Merton's model assumes that the obligor has sold the certain amount of bonds that do not pay any interest (zero-coupon bonds) at the certain date. The total debt $L$ must be repaid at the "maturity" date $T$ and default may happen only if $V(T) \leq L$. In reality, however, default may happen before the maturity of the debt. This situation has been first addressed in the Black-Cox extension of the Merton's model [5]. Black and Cox made two assumptions that are now common in the credit risk literature: i) company may default at any time $t \leq T$; ii) default happens at the *first passage*, i.e., irreversibly and instantly, whenever the diffusive path of $V(t)$ hits the absorbing default barrier, which is equal to $L$ or some lower threshold, $V_D$. Thus, calculation of the PoD was reduced to solution of the first stopping time problem: the process stops when the distance to the defaults boundary is zero. This problem has a long history in natural sciences [6] and has been frequently used in studies of extreme events in finance (see, e.g., [7] and references therein). Further development of structural models in the first-passage approximation includes non-zero coupon bonds [8], stochastic interest rates [9], and endogenously defined default boundaries [10].

It is well known, however, that the first-passage approximation may lead to underestimation of PoDs over short-time horizons and zero credit spreads for bonds with short time to maturity [11]. The credit spread is defined as a difference in yield between a risky debt instrument and the bond that is essentially credit risk-free, e.g., U.S. Treasury. It reflects the market price of potential credit accident and is always positive, albeit small close to maturity date. This discrepancy with economic reality is related to the deterministic nature of the Black-Cox-Merton framework, which employs a model of continues diffusion as an underlining transport mechanism of firm's assets value to the *absorbing* default boundary. As a result, in the first-passage approximation default is a predictable event. Lack of risk demands no premium and the theory predicts zero short-term credit spreads. To address this inconsistency Zhou introduced the structural model with discontinuous diffusion, allowing for sudden jumps in the firm's asset value [12]. This approach makes a credit event accidental and may explain positive short-term credit spreads.

Another class of default risk models is based on the phenomenological "reduced-form" approach [13 14 15]. In contrast to structural models, the actual cause of default is out of the modeling scope. Instead, it is assumed that the cumulative probability to survive default, $\Omega(t)$, obeys the following phenomenological kinetic equation [16]

$$\frac{d\Omega(t)}{dt} = -h(t)\Omega(t), \qquad \Omega(0) = 1. \qquad (1)$$

Here $h(t)$ is the "hazard rate" function, which represents the instantaneous probability of default at time $t$ conditional on the survival up to $t$. Note that this definition of $h(t)$ is applicable for both reduced-form and structural models [17] and allows for valuation of the market-implied cumulative PoD, $P(t) = 1 - \Omega(t)$, via Eq.(1). In reduced-form models the hazard rate for a firm is calibrated from market prices of various credit sensitive securities it has issued. Since there is always a market uncertainty regarding the default event, $h(t)$ is always positive and there is no issue with zero short-term credit spreads. This approach is self-contained, however, it is lacking the micro-economic interpretation of a default event, which is the key advantage of structural models.

Duffie and Lando were first to uncover the intrinsic connection between structural and reduced-form models of credit risk [18]. They consider the structural model with endogenously determined default barrier and postulated that investors observe noisy and delayed accounting reports. Since the key financial parameters are imperfectly observed, the firm's default incident is fundamentally unpredictable, conditional on the information available to the market observer [19]. The new methodology based on the notion of incomplete information provides the link between structural ("microscopic") and reduced-form ("macroscopic") models of credit risk and allows for *derivation* of the hazard rate, which involves certain basic assumptions regarding the cause of default [20 21 22 23 24 25 26 27]. In this context, it is remarkable that the kinetic equation formally equivalent to Eq.(1) has been derived almost a century ago by Smoluchowski [28]. He calculated the time-dependent intensity of an irreversible chemical reaction assisted by diffusion of reagents. The problem was solved within the first-passage approximation, which implies that recombination is accomplished at any contact. This model was extended by Collins and Kimball [29] who took into account that in reality not every encounter of reagents leads to reaction and introduced the intrinsic rate constant to characterize the efficiency of recombination at the contact. Obviously, introduction of this parameter reflects incomplete information regarding the elementary act of chemical transformation.

In this article, motivated by results obtained in physics and chemistry of complex systems, we formulate the extended Black-Cox model of default risk. We consider firm's ability to avoid default even if company's liabilities momentarily exceeding its assets. In other words, we assume that similarly to diffusion assisted irreversible reactions, not every contact with a default barrier inevitably leads to bankruptcy. Indeed, many financially distressed companies may adjust their liabilities and continue to service their debts. The uncertainty in these adjustments and/or other measures including possibility of a government intervention is not reflected by the traditional Black-Cox model and its extensions. Mathematically, we replace the absorbing boundary condition (BC) with the more flexible "radiation" (also called "reactive" or "elastic") BC. The latter accounts for a partial reflection as well as a diffusive flux across the boundary. Thereby, we naturally introduce incomplete information into the Black-Cox-Merton framework and make default unpredictable. The local rate of default at the boundary may be easily calibrated per credit rating's category, industry sector, country, and geographical region. Moreover, it can be used to reflect the "fine-structure" of the firm's financials, e.g., its size, cash holdings, debt liquidity premium, political environment, etc.

We formulate the problem of finding the cumulative corporate PoD and the relevant hazard rate in terms of the one-dimensional Fokker-Planck equation (FPE) with a linear potential and the radiation BC. The exact analytical solution of this equation allows for straightforward derivation of both the cumulative PoD and the relevant hazard rate. We also take into account uncertainty in the distance to the default barrier and demonstrate that the consequent generalization of the extended Black-Cox model leads to a positive credit spreads even for bonds with very short time to maturity. We show that the derived analytical expression for the real-world (physical) cumulative PoD fits the historical data on global corporate defaults in 1981 – 2008 [30] well. Remarkably, the straightforward re-parameterization of the cumulative PoD under the risk-neutral measure yields the hazard rate that qualitatively describes, without any additional fitting, the split behavior of credit spreads for bonds of companies in different categories of speculative-grade ratings with varying time to maturity. Introduction of the finite rate of default at the boundary improves valuation of credit risk for short time horizons, which is the key advantage of the extended Black-Cox model. We also consider the influence of uncertainty in the initial distance to default barrier on the outcome of the model and demonstrate that this additional source of incomplete information may be responsible for non-zero credit spreads for bonds with very short time to maturity.

**II. Diffusion in a constant field.**

Following the traditional approach, we begin with the simplest form of the Ito's stochastic differential equation that describes an irregular dynamics of the dimensionless distance to the default barrier, $x = \ln(V/L)$ [16]:

$$dx = a\, dt + \sigma\, dW. \tag{2}$$

Here $a = \mu - \sigma^2/2$ is the cumulative rate of expected growth of firm's assets and $W$ represents the standard Brownian motion (Wiener process). For simplicity, it is assumed that $V_D = L$ and the key parameters of the model: $\mu$ and $\sigma$ do not depend on time. In physics Eq.(2) may describe a strongly damped motion of Brownian particles in a gravitational field [31]. To proceed further it is convenient to reformulate the problem in terms of the one-dimensional FPE. From standard results in the probability theory (see, e.g., [32]), the conditional probability $p(x,t|x_0,0) \equiv p(x,t \mid x_0)$ of finding the distance to the default boundary equal to $x$ at the end of the interval $(\tau = 0, t]$, if $x$ is known to has been equal to $x_0$ at its beginning, satisfies the following FPE

$$\frac{\partial p(x,t|x_0)}{\partial t} = -\frac{\partial J(x,t|x_0)}{\partial x}. \tag{3}$$

Here we define the probability density flux at any point $x$ as [32]

$$J(x,t|x_0) = a\, p(x,t|x_0) - D\frac{\partial p(x,t|x_0)}{\partial x}. \tag{4}$$

The FPE Eq.(3) has the familiar form of the local conservation law. It describes diffusion with the constant coefficient $D = \sigma^2/2$ in a static linear potential, which is responsible for the drift with the constant coefficient $a$. Thus, with the corresponding re-definition of the parameters, the time evolution of the distance to default barrier is equivalent to, e.g., adsorption of Brownian particles in a liquid under the influence of a gravitation field or motion of electrons in amorphous insulators under the influence of electric field. The FPE Eq.(3) is usually solved with the following conditions

$$p(x,0|x_0) = \delta(x - x_0) \; , \tag{5}$$

$$p(x,t|x_0)|_{x \to \infty} = 0 \; . \tag{6}$$

If the conditional probability $p(x,t|x_0)$ is known, the cumulative PoD and the relevant hazard rate can be straightforwardly calculated:

$$P(t|x_0) = 1 - \int_{0+}^{\infty} p(x,t|x_0)dx = 1 - \Omega(t|x_0), \tag{7}$$

$$h(t|x_0) = -\Omega^{-1}(t|x_0)\frac{d\Omega(t|x_0)}{dt}. \tag{8}$$

Notice that, as follows from Eqs.(3), (6) – (8), the time derivative of the cumulative survival probability is equal to the probability density flux at the boundary

$$\frac{d\Omega(t|x_0)}{dt} = J(x,t|x_0)|_{x \to 0} = -\frac{dP(t|x_0)}{dt} \; . \tag{9}$$

In order to determine the unique solution $p(x,t|x_0)$ we must specify the relevant BC. The first-passage approximation is equivalent to the absorbing BC, which implies the vanishing density at the boundary, $p(0,t | x_0) = 0$. Therefore, as follows from Eq.(4), the absorbing BC can be represented as

$$J(x,t | x_0)\big|_{x \to 0} = -D \frac{\partial p(x,t | x_0)}{\partial x}\bigg|_{x \to 0}. \tag{10}$$

In the limit of small cumulative PoD, $P(t|x_0) \ll 1$, the survival probability is high, $\Omega(t|x_0) \sim 1$, and Eq.(8) can be reduced to

$$h(t | x_0) \approx -\frac{d\Omega(t | x_0)}{dt} = \frac{dP(t | x_0)}{dt}. \tag{11}$$

This approximation in concert with the general formula Eq.(9) and the absorbing BC Eq.(10) immediately leads to the following expression for the hazard rate

$$h(t | x_0) \approx D \frac{\partial p(x,t|x_0)}{\partial x}\bigg|_{x \to 0}, \tag{12}$$

which was first obtained by probabilistic methods by Duffie and Lando in Ref.[19].

The solution $p(x,t | x_0)$ of Eq.(3) that satisfies the initial condition Eq.(5) is known as Green's function. For the absorbing BC this function is well-known [33]:

$$p(x,t | x_0) = \frac{1}{2\sqrt{\pi D t}} \exp\left[\frac{a(x - x_0)}{2D} - \frac{a^2 t}{4D}\right] \left\{ \exp\left[-\frac{(x - x_0)^2}{4Dt}\right] - \exp\left[-\frac{(x + x_0)^2}{4Dt}\right] \right\}. \tag{13}$$

Substitution of this expression into Eq.(7) after integration over *x* yields the Black-Cox result for the cumulative PoD

$$P(t|x_0) = \Phi\left[-\frac{x_0 + at}{\sqrt{2Dt}}\right] + \exp(-ax_0/D)\Phi\left[-\frac{x_0 - at}{\sqrt{2Dt}}\right]. \tag{14}$$

Here $\Phi(z) = \frac{1}{2}(1 + erf\frac{z}{\sqrt{2}})$ is the cumulative normal distribution and $erf(z)$ is the error function. It follows from Eqs.(7), (8) and (14) that within the first-passage approximation the hazard rate acquires the following "microscopic" definition

$$h(t \mid x_0) = \frac{x_0(4\pi Dt^3)^{-1/2}\exp[-(x_0 + at)^2/4Dt]}{\Phi\left[\frac{x_0 + at}{\sqrt{2Dt}}\right] - \exp(-ax_0/D)\Phi\left[-\frac{x_0 - at}{\sqrt{2Dt}}\right]}. \tag{15}$$

The numerator of this expression is the time derivative of the cumulative PoD or the density of the first hitting-time [34].

Taking into account that $erf(z) \sim 1 - \frac{\exp(-z^2)}{\sqrt{\pi z}}[1 + O(1/z^2)]$ and $erf(-z) = -erf(z)$, it is easy to see that for bonds with short time to maturity, $t \ll x_0/|a|$, the cumulative PoD, Eq.(14), and the hazard rate, Eq.(15), are going to zero extremely fast, which contradicts economic reality. Another erroneous forecast of the basic Black-Cox model is readily seen from the following example. Let us assume that initially the firm's assets value is equal to the value of its liabilities $V(0) = L$ ($x_0 = 0$). In this case, Eq.(14) predicts that the company has no chance to escape a default, $P(t|0) = 1$. On the other hand, for any $x_0 > 0$, it forecasts that $P(0|x_0 > 0) = 0$. Apparently, this discontinuity is the direct consequence of the absorbing BC, which implies the vanishing probability density at the boundary.

In what follows we relax the first-passage approximation by replacing the absorbing BC with the radiation BC and explore the results of this operation. It has been already mentioned in the Introduction that the radiation BC allows for a partial reflection from the default barrier as well as the flux across the boundary. Since the latter is proportional to the probability density at the barrier, the radiation BC takes the following form [29]

$$J(x,t \mid x_0)|_{x \to 0} = -k_c p(0,t|x_0), \qquad (16)$$

where $k_c$ describes a local rate-constant of default at the boundary. In essence, Eq.(16) reflects the likelihood for financially distressed company that arrives at the barrier in a predictable fashion to escape a default incident. The probability to survive a contact with the default barrier is negatively correlated with $k_c$. If it is zero, the BC becomes purely reflective $J(x,t \mid x_0)|_{x \to 0} = 0$. Apparently, the first-passage approximation, Eq.(10), corresponds to the opposite limit, $k_c \to \infty$. Similarly to Collins-Kimball extension of the basic Smoluchowski model, the assumption of the finite $k_c$ introduces incomplete information into the traditional Black-Cox model and makes a default event unpredictable.

The exact Green's function for the FPE Eq.(3) with the radiation BC is known [33]:

$$\begin{aligned} p(x,t \mid x_0) = \frac{1}{2\sqrt{\pi Dt}} & \left\{ \exp\left[-\frac{(x-x_0-at)^2}{4Dt}\right] + \exp\left[-\frac{(x+x_0-at)^2}{4Dt} - \frac{ax_0}{D}\right] \right\} \\ & - \frac{(a+2k_c)}{D} \exp\left[\frac{(a+k_c)(x+k_c t) + k_c x_0}{D}\right] \Phi\left[-\frac{(a+2k_c)t + x + x_0}{\sqrt{2Dt}}\right] \end{aligned} \qquad (17)$$

This expression provides the complete solution of the problem, i.e., the entire conditional probability function that satisfies the boundary condition Eq.(16) as well as usual conditions Eqs.(5) and (6). Substitution of this expression into Eq.(7) leads to

$$P(t|x_0) = \Phi\left[-\frac{x_0 + at}{\sqrt{2Dt}}\right] + \frac{k_c}{k_c + a}\exp(-ax_0/D)\Phi\left[-\frac{x_0 - at}{\sqrt{2Dt}}\right]$$
$$- \frac{2k_c + a}{k_c + a}\exp\{[x_0 + (k_c + a)t]\frac{k_c}{D}\}\Phi\left[-\frac{x_0 + (a + 2k_c)t}{\sqrt{2Dt}}\right]. \quad (18)$$

To determine the hazard rate function we need the time-derivative of the $P(t|x_0)$, which can be obtained either directly from Eq.(18) or deduced from Eq.(9) and the radiation BC, Eq.(16). Both approaches yield the same result

$$\dot{P}(t|x_0) = -\dot{\Omega}(t|x_0) = k_c\left\{\frac{1}{\sqrt{\pi Dt}}\exp[-\frac{(x_0 + at)^2}{4Dt}] - \right.$$
$$\left. -\frac{2k_c + a}{D}\exp\{[x_0 + (k_c + a)t]\frac{k_c}{D}\}\Phi\left[-\frac{x_0 + (a + 2k_c)t}{\sqrt{2Dt}}\right]\right\} \quad (19)$$

which is valid at any rate of default at the boundary.

It is easy to see from Eqs.(18) and (19) that in the extended Black-Cox model the cumulative PoD and the hazard rate are determined by the interplay between the rate of diffusion, i.e., volatility of asset value, the drift of firm's assets, and the local rate of default at the boundary. The complex PoD's kinetics described by Eq.(18) is far from exponential and differs from predictions of the traditional Black-Cox model. As expected, for large default rates $k_c >> |a|$, the first two terms on the right-hand side of Eq.(18) reproduce the Black-Cox result, Eq.(14). However, since not every contact with the radiation boundary leads to default, the cumulative PoD described by Eq.(18) is generally smaller than the one predicted by Eq.(14). Note that the third term, which is responsible for this decrease disappears only if $k_c \to \infty$, which corresponds to the first passage approximation. In the opposite limit, $k_c = 0$, Eq.(18) yields $P(t|x_0) = 0$ as it should be in the toy-economy with no defaults. The cumulative PoD described by Eq.(18)

is well defined even in the special case $k_c + a = 0$ ($a < 0$) [33]. In general, for $a < 0$, the drift is directed toward the default barrier and the firm's default is inevitable $P(t|x_0) \to 1$ as $t \to \infty$. On the contrary, if $a > 0$, Eq.(18) predicts the universal long-time asymptotic

$$P(\infty | x_0) = \frac{k_c}{k_c + a} \exp(-ax_0/D). \tag{20}$$

From the physical point of view this result reflects the competition between sink at the boundary and escape to infinity (drift away from the boundary). It corresponds to the steady state balance between these two processes that is established at long times. The physical interpretation of Eq.(20) is that diffusion against the flow is possible. However, the stationary probability for Brownian particle to reach the boundary for any initial distance is exponentially small. The pre-exponential factor $k_c/(k_c + a)$ in Eq.(20) simply tells us that, contrary to the first-passage approximation, the system has a chance to survive even if its trajectory hits or starts at the boundary. Let us take a closer look at the latter case. Assume that initially the firm's assets value is equal to the value of its liabilities, $x_0 = 0$. In this case, the Black-Cox model, Eq.(14) always yields $P(t|0) = 1$, whereas Eq.(18) in the limit of pure diffusion, $a = 0$, gives[35]

$$P(t|0) = 1 - \exp(t/t_0) \, erfc(\sqrt{t/t_0}) = \begin{cases} 2\sqrt{t/\pi t_0}, & t \ll t_0 \\ 1 - \sqrt{t_0/\pi t}, & t \gg t_0 \end{cases}. \tag{21}$$

Here $erfc(z) = 1 - erf(z)$ is the complementary error function and we introduce the characteristic time of the process $t_0 = D/k_c^2$. Thus, even in this extreme case, the extended Black-Cox model forecasts the power law for the time-dependence of the cumulative PoD. Remarkably, in the short-term region, $t \ll t_0$, the PoD decreases to zero as $\sim k_c \sqrt{t}$, i.e., much slower than in the first-passage approximation. We come back to this important result in Section IV. Note that since Eq.(21) describes the situation with the starting point at the default barrier and $a = 0$, it predicts an increase in the cumulative PoD with a decrease in the volatility of $V$, $P(t|0) \sim 1/\sqrt{Dt} \sim 1/(\sigma\sqrt{t})$, in the short-term region.

In general, for long times or small volatilities and large default rates at the boundary, $t \gg t_0$, the time derivative of the cumulative PoD, Eq.(19), simplifies to

$$\frac{dP(t|x_0)}{dt} = \frac{k_c x_0}{(at + 2k_c t + x_0)\sqrt{\pi Dt}} \exp\left[-\frac{(x_0 + at)^2}{4Dt}\right] \qquad (22)$$

and further reduces to the following asymptotic

$$\frac{dP(t|x_0)}{dt} = \frac{x_0}{2\sqrt{\pi D}\, t^{3/2}} \exp\left[-\frac{(x_0 + at)^2}{4Dt}\right] \qquad (23)$$

when $k_c \to \infty$. As expected, this expression coincides with the time derivative of the cumulative PoD obtained in the first-passage approximation [cf. Eq.(15)].

Although in the extended Black-Cox model a default incident is unpredictable, it still requires some time for continues diffusion to reach the barrier. Therefore, it is not surprising that, as follows from Eqs.(8), (18), and (19), in the limit $Dt \to 0$ the hazard rate is determined by the following general expression

$$h(t|x_0)|_{Dt \to 0} = k_c \delta(x_0 + 0_+). \qquad (24)$$

It is easy to see from Eqs.(18) and (19) that for large initial distance to the default barrier, $t \ll x_0/|a|$, the cumulative PoD and the hazard rate are going to zero very fast, which may lead to severe underestimation of a default risk for bonds of companies in investment grade ratings categories ('BBB' or better). This behavior does not depend on the choice of BC and should be expected in any continues diffusion model with known distance to the default barrier and initial condition fixed by Eq.(5). This issue may be addressed only by consideration of other possible causes of incomplete information, e.g., uncertainty in the default barrier, time-shift (information lag) in the Brownian motion, etc. In the next section we consider the influence of uncertainty in the initial condition on the outcome of the extended Black-Cox model.

### III. Uncertainty in the Distance to Default Barrier.

In our study we assume that the radiation BC remains the same and the initial condition acquires the following form

$$p_\delta(x,0\,|\,x_0) = \frac{1}{\delta\sqrt{2\pi}} \exp\left[-\frac{(x-x_0)^2}{2\delta^2}\right]. \tag{25}$$

In other words, we suppose that the initial value of $x$ is normally distributed with the mean $<x_0> = x_0$ and variance $<(x-x_0)^2> = \delta^2$. In general, the exact solution of the FPE is described by the convolution of its Green's function with an arbitrary initial condition. Therefore, in the case under consideration, the exact solution of the problem is determined by the following quadrature

$$p_\delta(x,t\,|\,x_0) = \int_0^\infty p(x,t\,|\,y)\,p_\delta(y,0\,|\,x_0)\,dy, \tag{26}$$

where $p(x,t|x_0)$ is defined by Eq.(17). In what follows we concentrate on the short-term limit and obtain analytical expressions for the cumulative PoD and the hazard rate.

The peak of the initial distribution $p_\delta(x,0|x_0)$ at $x = x_0$ moves away ($a > 0$) from or towards ($a < 0$) the default barrier and expands due to diffusion as $x(t) \sim \sqrt{Dt}$. For short time periods diffusion dominates, $at << \sqrt{Dt}$. The diffusion length $\sqrt{Dt}$ provides a measure of how far the density has propagated in the *x*-direction by diffusion at the instance *t*. At short times determined by the inequality $\sqrt{Dt} << x_0$ Eq.(17) reduces to

$$p(x,t|x_0) \approx \frac{1}{2\sqrt{\pi Dt}} \exp\left[-\frac{(x-x_0-at)^2}{4Dt}\right]. \qquad (27)$$

Comparing this expression with Eq.(25) we obtain, up to the normalization constant,

$$p(x,\tau|x_0 - a\tau) \approx p_\delta(x,0|x_0), \qquad (28)$$

$$\tau = \frac{\delta^2}{2D}. \qquad (29)$$

Thus, as long as $\sqrt{Dt} << x_0$, the randomness in $x_0$ is translated into the space- and time-shift in the Green's function of Eq.(3). Further, it is not difficult to see that in this limit

$$p(x,t+\tau|x_0 - a\tau) = p_\delta(x,t|x_0)\Omega(\tau|x_0 - a\tau), \qquad (30)$$

where $\Omega(\tau|x_0 - a\tau)$ is the probability to survive default starting diffusion at $x_0 - a\tau$ by the time $\tau$. After integration over *x*, this expression yields

$$\Omega_\delta(t|x_0) = \Omega(t+\tau|x_0 - a\tau)/\Omega(\tau|x_0 - a\tau), \qquad (31)$$

which leads to the generalized formulas for the short-term cumulative PoD and the hazard rate:

$$P_\delta(t|x_0) = 1 - \Omega_\delta(t|x_0) = \frac{P(t+\tau|x_0 - a\tau) - P(\tau|x_0 - a\tau)}{\Omega(\tau|x_0 - a\tau)}, \qquad (32)$$

$$h_\delta(t \mid x_0) = -\frac{d\Omega(t + \tau \mid x_0 - a\tau)}{dt} \Omega^{-1}(t + \tau \mid x_0 - a\tau), \tag{33}$$

where

$$\Omega(\tau \mid x_0 - a\tau) = \Phi\left[\frac{x_0}{\sqrt{2D\tau}}\right] - \frac{k_c}{k_c + a} \exp[-\frac{a}{D}(x_0 - a\tau)]\Phi\left[-\frac{x_0 - 2a\tau}{\sqrt{2D\tau}}\right] \\ + \frac{2k_c + a}{k_c + a} \exp[(x_0 + k_c\tau)\frac{k_c}{D}]\Phi\left[-\frac{x_0 + 2k_c\tau}{\sqrt{2D\tau}}\right] . \tag{34}$$

It is naturally to assume that the uncertainty in the initial value of $x$ is less than its mean value. Therefore, $a\tau < \sqrt{D\tau} \ll x_0$ and $\Omega(\tau \mid x_0 - a\tau) \approx 1$. However, this term guarantees the proper normalization of the solution, Eq.(30), and we should keep $P(\tau \mid x_0 - a\tau)$ in the numerator of Eq.(32). If, for example, one would follow the CreditGrades approach and assume that $\Omega(\tau \mid x_0 - a\tau) = 1$ (or equivalently $P(\tau \mid x_0 - a\tau) = 0$), then in the first-passage approximation Eq.(32) reduces to the formula for the cumulative survival probability obtained in Ref.[20]:

$$\Omega_\delta(t \mid x_0) = \Phi\left[-\frac{A}{2} + \frac{x_0 + \delta^2/2}{A}\right] - \exp(\tilde{x}_0)\Phi\left[-\frac{A}{2} - \frac{x_0 + \delta^2/2}{A}\right], \tag{35}$$

where $A = \sqrt{2D\tilde{t}}$, $\tilde{t} = t + \tau$ [36]. This expression, however, yields $\Omega_\delta(0 \mid x_0) < 1$, which contrary to Eq.(32), implies a non-zero probability of default at $t = 0$. The reason for this erroneous behavior is the omitted $\Omega(\tau \mid x_0 - a\tau)$ term in Eq.(30).

Evidently, the assumption of normally distributed initial value of $x$ leads to positive hazard rates at short times, which implies non-zero credit spreads even for bonds with very short time to maturity. At $t = 0$ Eqs. (32) – (34) in concert with Eqs.(18) and (19) yield

$$P_\delta(0 \mid x_0) = 0, \tag{36}$$

$$h_\delta(0 \mid x_0) = \frac{k_c}{\Omega(\tau \mid x_0 - a\tau)} \left\{ \frac{\exp(-x_0^2/4D\tau)}{\sqrt{\pi D\tau}} - \frac{2k_c + a}{D} \exp[(x_0 + k_c\tau)\frac{k_c}{D}] \Phi\left[-\frac{x_0 + 2k_c\tau}{\sqrt{2D\tau}}\right] \right\}$$

$$\approx \sqrt{\frac{2}{\pi}} \frac{k_c}{\delta} \exp(-\frac{x_0^2}{2\delta^2}) \tag{37}$$

As expected, the value of $h_\delta(0 \mid x_0)$ strongly depends on the ratio $x_0/\delta$. Notably, it is completely determined by $k_c$, $x_0$, $\delta$ and does not depend on the transport to the default boundary, i.e., volatility of firm's assets. In conclusion of this section, we would like to stress that randomness of the initial value of $x$ may be considered as a good proxy for uncertainty in the endogenously defined default barrier and vice versa. This approximation is broadly used by the CreditGrades model, which was jointly developed by JP Morgan, Goldman & Sachs, Deutsche Bank and RiskMetrics (see Ref. [20]).

**IV. Historical PoD Data and the Term-Structure of Credit Spreads.**

We begin this section with a comparison of the observed term-structure (kinetics) of the global cumulative PoD [30] with the theoretical predictions, Eqs.(14) and (18). In spite of the long-term observations, 1981 – 2008, some historical data on cumulative defaults have low statistical significance, especially in the investment grade category. For example, throughout the 28-years of observations, only five companies originally rated 'AAA' by credit agencies have ever defaulted. Therefore we limit our consideration by issuers that were originally rated by Standard & Poor's in the 'BB' and 'B' categories. Since 1981, these rating categories have accounted for 431 defaulters (25.8% of the total) and 897 defaulters (53.8% of the total), correspondingly.

To reduce the number of fitting parameters we assume that $\delta = 0$ and for convenience of a reader replace $D$ with $\sigma^2/2$ in the formula for the term structure of the real-world (*physical*) cumulative PoD, Eq.(18), which we present here in a form more familiar in the credit risk literature:

$$P(t|x_0) = \Phi\left[-\frac{x_0 + at}{\sigma\sqrt{t}}\right] + \frac{k_c}{k_c + a}\exp(-2ax_0/\sigma^2)\Phi\left[\frac{-x_0 + at}{\sigma\sqrt{t}}\right] \\ - \frac{2k_c + a}{k_c + a}\exp\{[x_0 + (k_c + a)t]\frac{2k_c}{\sigma^2}\}\Phi\left[-\frac{x_0 + (a + 2k_c)t}{\sigma\sqrt{t}}\right] \quad . \quad (38)$$

The historical PoD-data in the 'BB' and 'B' categories are presented in Table 1. These values represent the change in time of the observed PoD following the *initial* rating of the firm [30]. We utilized the full available dataset and performed the standard stochastic optimization of the parameters [37] for the root mean square deviation (RMSD) $\rho$ with the objective function:

$$\varepsilon(\vec{Z}) = \sqrt{N^{-1}\sum_{i=1}^{n}\omega_i\left[P(t_i,\vec{Z}) - P_i^h\right]^2} \quad . \quad (39)$$

Here $P_i^h$ is the historical PoD at $t = t_i$, $\omega_i$ represents the relevant weight, $\vec{Z}$ is the vector of parameters, and $N = \sum_{i=1}^{n}\omega_i$ is the normalization factor. We have used a random search algorithm with uniform distribution in the interval $[z_n q, z_n/q]$, where $z_n$ is the n-th parameter from the previous set of parameters that lowered $\rho$ [37]. The initial set of parameters was obtained from the preliminary fit in Excel. The number of trials was chosen $10^4$ and q = 0.8.

The fitted cumulative PoDs, the relevant RMSDs, and the corresponding normalized parameters $\tilde{a} = a/\sigma$, $\tilde{x}_0 = x_0/\sigma$, $\tilde{k}_c = k_c/\sigma$ resulting from the fitting procedure are shown in Tables 1 and 2. In Figure 1a we plot the calculated and observed global corporate cumulative PoDs (1981-2008) [30]. Our results demonstrate that both 2-parameter Black-Cox model and 3-parameter extended Black-Cox model fit well the historical data on global corporate defaults in the 'BB' and 'B' ratings categories. Not surprisingly, the RMSDs for 3-parameter extended Black-Cox model are better. However, the differences in the calculated values of the PoD are basically insignificant for time horizon more than three years. It is clearly seen that the difference in the firm's initial distance to default barrier measured in standard deviations (see Table 2) largely determines its default behavior over the extended period of time. Notably, Lei and Hawkins came to the same conclusion within the framework of the classical Black-Cox model [38]. These authors analyzed data on the time evolution of the cumulative PoD and its dependency on the originally assigned credit rating over the 15 years published by Standard and Poor's in 2000 [39]. Interestingly, the new data published by Standard and Poor's in 2009 [30] did not considerably change estimated initial distances to default for an average 'BB' and 'B' company (~ 3 and ~ 2 standard deviations from the default boundary, correspondingly) made by Lei and Hawkins. It should be clear from the above that the extended Black-Cox model leads to generally smaller values of $\tilde{x}_0$ (~ 2 and ~ 1 standard deviations from the default boundary for an average 'BB' and 'B' company) than those calculated by these authors. Note that both models lead to the normalized drift, $\tilde{a}$, that is positive and quite similar for an average 'BB' and 'B' company (see Table 2 and Ref.[38]). The key difference between the absorbing and radiation BCs is clearly seen in Figure 1b, which presents the kinetics of the cumulative PoD in the short-time region. It has been shown in Section II that the extended Black-Cox model forecasts the relatively slow, $\sim k_c \sqrt{t}$, decrease in the cumulative PoD in the short-term region, $t << t_0 = 0.5 \tilde{k}_c^{-2}$ ($t \sim 1 \div 3$ years, see Table 2). In other words, in the short-term the risk of default disappears slower in the model with the radiation BC than in the first-passage approximation.

It is well known that the continuously compound credit spread for a zero-recovery, non-callable, zero-coupon bond is completely determined by the hazard rate [16]. Therefore, as follows from Eqs.(7) and (8), the term-structure of the credit spread $s(t | x_0)$ is determined by the following expression

$$s(t | x_0) = h(t | x_0) = \frac{\dot{P}(t | x_0)}{1 - P(t | x_0)} \qquad . \qquad (40)$$

Being proportional to the time-derivative of the cumulative PoD, it is very sensitive to the details of its changes in time. The term-structure of credit spreads may drastically change depending on the time left to maturity of the debt and can have upward, humped, and downward shapes. Empirical studies demonstrate a split behavior of credit spreads in different credit rating categories as time to the maturity of the risky bond varies [40 41 42]. For instance, bonds issued by companies in the investment-grade category tend to have spreads that are slowly growing with maturity, which has been explained by a growth of the PoD over a long period of time. On the other hand, issuers with higher credit risk (lower rating) tend to have humped and decreasing spreads with maturity. This, at the first glance, counterintuitive pattern was qualitatively explained by the "crisis-at-maturity" model [42], which emphasized the difficulties that companies in the speculative-grade category may come across with refinancing as their short-term bonds matures. Consequently, the PoD in the short term may be higher than in the long run.

What really matters, however, is the rate of change in the cumulative PoD with the variation in time to maturity (see Eq.(40)).

For simplicity we assume that the expected return rate is equal to the risk-free interest rate, i.e., in the definition of the parameter $a$ we set $\mu = r$. Hence, we ignore the difference between the "actuarial" and actual market spreads [16]. In this case, the analytical expressions for the cumulative PoD, Eq.(38), its time derivative

$$\dot{P}(t|x_0) = k_c \left\{ \frac{1}{\sigma\sqrt{\pi t/2}} \exp[-\frac{(x_0 + at)^2}{2\sigma^2 t}] - \right.$$
$$\left. - 2\frac{a + 2k_c}{\sigma^2} \exp\{[x_0 + (k_c + a)t]\frac{2k_c}{\sigma^2}\} \Phi\left[-\frac{x_0 + (a + 2k_c)t}{\sigma\sqrt{t}}\right] \right\} \quad (41)$$

and Eq.(40) enable valuation of the term-structure of credit spreads in the extended Black-Cox model. Formula (15) is used to calculate the same in the standard Black-Cox model. In Figure 2 we plot the calculated kinetics of credit spreads for zero-coupon, zero-recovery bonds. We use the same set of parameters resulting from the excellent fitting of historical cumulative PoDs in Figure1a. Qualitatively, charts are in line with empirical shapes of the term-structure of credit spreads and reflect the fact that the lower the credit rating of an issuer, the higher is the credit spread [40 – 42]. Both models forecast the split behavior with varying time to maturity for bonds of companies in different categories of speculative-grade credit ratings. More precisely, charts on Figure 2 qualitatively illustrate that as the company's financial situation improves, the short-term credit spread curves may change dramatically: from the humped (for 'B'-rated firms) to the upward shape (for 'BB'-rated firms). Quantitatively, however, the differences in the calculated values of credit spreads between these two models are significant for short-term bonds. It is clearly seen from Figure 2 that introduction of the finite rate of default at the boundary may improve valuation of credit spreads for short-maturity debt. This is one of the main advantages of the extended Black-Cox model.

The drawback of both models - decline in the value of credit spreads in the short-term region - is also evident from the theoretical curves on Figure 2. Therefore, in Figures 3a and 3b we plot the kinetics of credit spreads predicted by the extended Black-Cox model for varying levels of market's uncertainty in the perceived initial distance to the default barrier, see Eqs.(32) - (34). Different curves in Figures 3a and 3b correspond to different levels of normalized uncertainty, $\tilde{\delta} = \delta/\sigma$, for 'B'- and 'BB'-rated companies. Naturally, we keep parameters $\tilde{a}$, $\tilde{x}_0$, and $\tilde{k}_c$ the same as resulting from the fitting of observed global cumulative PoD (see Table 2). These figures clearly demonstrate that account of uncertainty in the firm's initial distance to the default barrier may lead to non-zero spreads for bonds with very short time to maturity (less than three months).

**V. Conclusion**

This paper develops a straightforward generalization of the Black-Cox structural model of default risk. Diffusion in a linear potential with the radiation BC is used to mimic a company's default process. The extended model captures uncertainty related to a chance to avoid default by firms with liabilities momentarily exceeding their assets. Finite local rates of default at the boundary may reflect the "fine-structure" of the firm's financials, e.g., its size, cash holdings and liquidity, or the political environment. We point out that the line between complete and incomplete information models can be drawn at the default boundary in terms of the relevant BC. Introduction of the local rate of default at the boundary elevates the limitations of the first-passage approximation and naturally introduces incomplete information into the Black-Cox-Merton deterministic framework.

We obtain the analytical expression for the real-world cumulative PoD that fit well the historical data on global cumulative corporate defaults. We also derive the analytical expression for the relevant hazard rate and demonstrate that under the risk-neutral measure both diffusion-based structural models - with absorbing and radiation BCs - may qualitatively explain the split behavior of credit spreads for bonds of 'BB'- and 'B'-rated companies with varying time to maturity. The radiation BC leads to the relatively slow, $\sim k_c \sqrt{t}$, decrease in the cumulative PoD in the short-term region, $t << 0.5 \tilde{k}_c^{-2}$. Consequently, it yields higher credit risk and significantly broader credit spreads for short-maturity debt than the first-passage approximation. This is the key advantage of the extended Black-Cox model.

Finally, we consider the influence of uncertainty in the distance to the default barrier on the outcome of the extended Black-Cox model and demonstrate that this additional source of incomplete information may be responsible for non-zero credit spreads for bonds with very short time to maturity (less than tree months). Left for future research is application of the extended Black-Cox model in the multi-firm scenarios for portfolio risk valuations.

**Acknowledgement**

We are extremely grateful to Anatoly I. Burshtein for many enlightening discussions on a range of kinetic problems dealt with in this article. This paper has not been prepared by YAK and NVS in their capacity as employees of any organization and reflects only their own views.

**Table 1.** Global cumulative corporate PoD (%) in ratings categories 'BB' and 'B' as reported in Ref.[30] and obtained by fitting the Black-Cox and extended Black-Cox models, Eqs.(14) and (38), correspondingly.

| Year | Observed PoD ('BB') | Black-Cox | Extended Black-Cox | Observed PoD ('B') | Black-Cox | Extended Black-Cox |
|---|---|---|---|---|---|---|
| 1 | 0.99 | 0.21 | 0.77 | 4.51 | 2.37 | 4.50 |
| 2 | 2.88 | 2.10 | 3.07 | 9.87 | 8.71 | 10.14 |
| 3 | 5.07 | 4.71 | 5.42 | 14.43 | 13.94 | 14.40 |
| 4 | 7.18 | 7.17 | 7.51 | 17.97 | 17.88 | 17.69 |
| 5 | 9.07 | 9.31 | 9.34 | 20.58 | 20.90 | 20.34 |
| 6 | 10.90 | 11.12 | 10.94 | 22.67 | 23.27 | 22.53 |
| 7 | 12.41 | 12.65 | 12.34 | 24.46 | 25.18 | 24.38 |
| 8 | 13.74 | 13.96 | 13.59 | 25.93 | 26.75 | 25.96 |
| 9 | 15.00 | 15.08 | 14.70 | 27.17 | 28.06 | 27.34 |
| 10 | 16.02 | 16.05 | 15.70 | 28.41 | 29.17 | 28.55 |
| 11 | 16.89 | 16.90 | 16.61 | 29.54 | 30.13 | 29.62 |
| 12 | 17.64 | 17.64 | 17.43 | 30.50 | 30.95 | 30.58 |
| 13 | 18.28 | 18.29 | 18.18 | 31.45 | 31.67 | 31.45 |
| 14 | 18.77 | 18.87 | 18.87 | 32.32 | 32.30 | 32.23 |
| 15 | 19.33 | 19.39 | 19.50 | 33.15 | 32.86 | 32.95 |
| 16 | 19.87 | 19.85 | 20.08 | 33.78 | 33.36 | 33.60 |
| 17 | 20.40 | 20.26 | 20.63 | 34.28 | 33.81 | 34.20 |
| 18 | 20.98 | 20.64 | 21.13 | 34.79 | 34.21 | 34.76 |
| 19 | 21.81 | 20.98 | 21.608 | 35.25 | 34.57 | 35.27 |
| 20 | 22.96 | 21.28 | 22.04 | 35.57 | 34.90 | 35.75 |

**Table 2**. The relevant RMSDs and calculated fitting parameters.

| | Black-Cox 'B' | Extended Black-Cox 'B' | Black-Cox 'BB' | Extended Black-Cox 'BB' |
|---|---|---|---|---|
| $\rho =$ | 0.75 | 0.14 | 0.31 | 0.22 |
| $\tilde{x}_0 =$ | 2.07 | 1.09 | 2.86 | 1.72 |
| $\tilde{a} =$ | 0.23 | 0.14 | 0.24 | 0.15 |
| $\tilde{k}_c =$ | n/a | 0.25 | n/a | 0.18 |

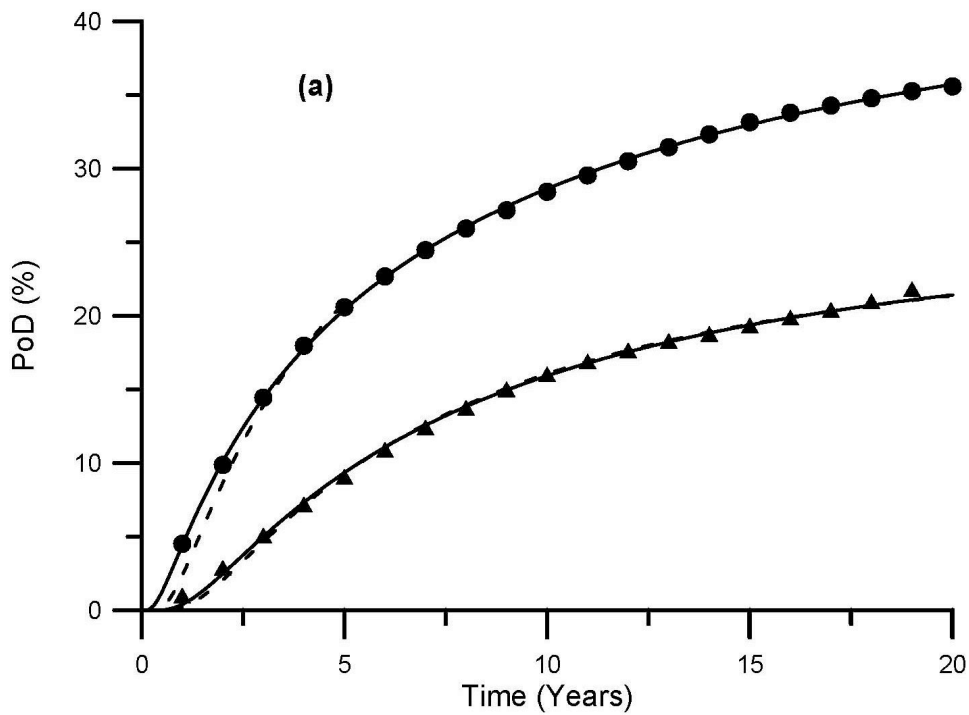

**Figure 1. Observed and calculated global corporate cumulative PoDs (1981-2008).**
Circles – ratings category 'B', triangles – ratings category 'BB'. Solid lines – extended Black-Cox model, dashed lines – standard Black-Cox model. For the relevant data-points see Table 1. Fitting parameters are shown in Table 2.

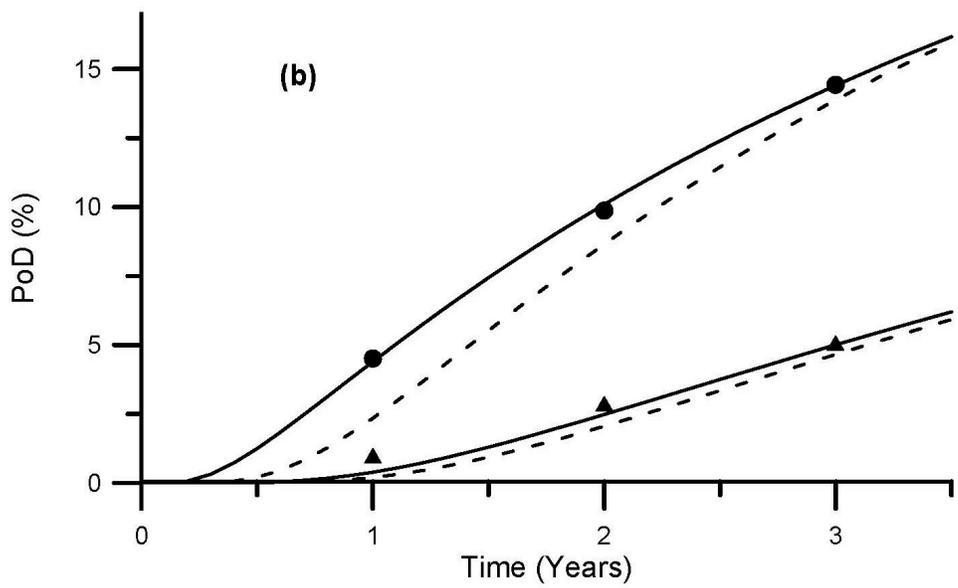

**Figure 1. Observed and calculated global corporate cumulative PoDs (1981-1984).** Circles – ratings category 'B', triangles – ratings category 'BB'. Solid lines – extended Black-Cox model, dashed lines – standard Black-Cox model. For the relevant data-points see Table 1. Fitting parameters are shown in Table 2.

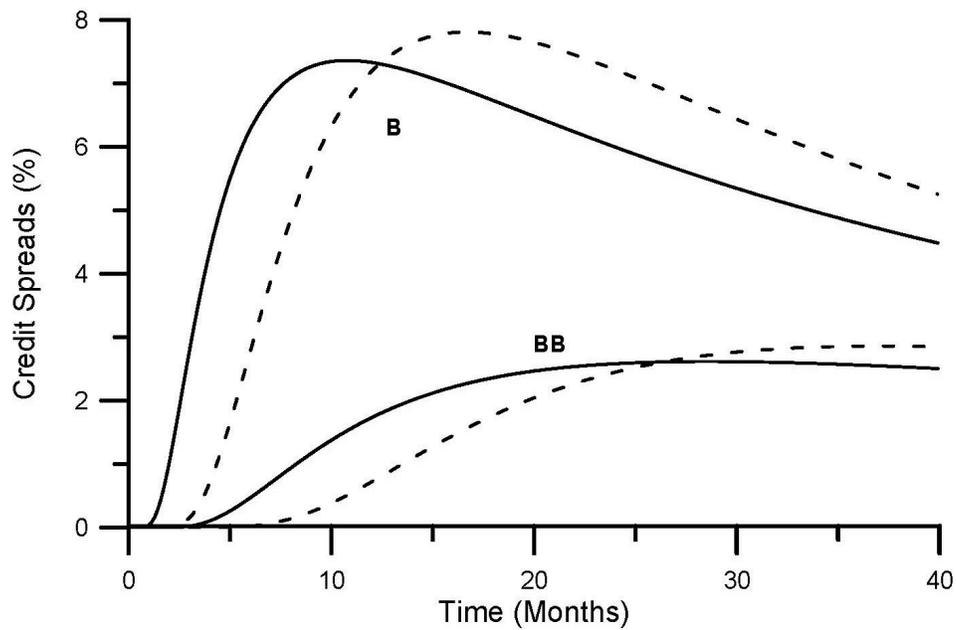

**Figure 2. Calculated kinetics (term-structure) of credit spreads.**

Solid lines – extended Black-Cox model, dashed lines – standard Black-Cox model.

Fitting parameters are shown in Table 2.

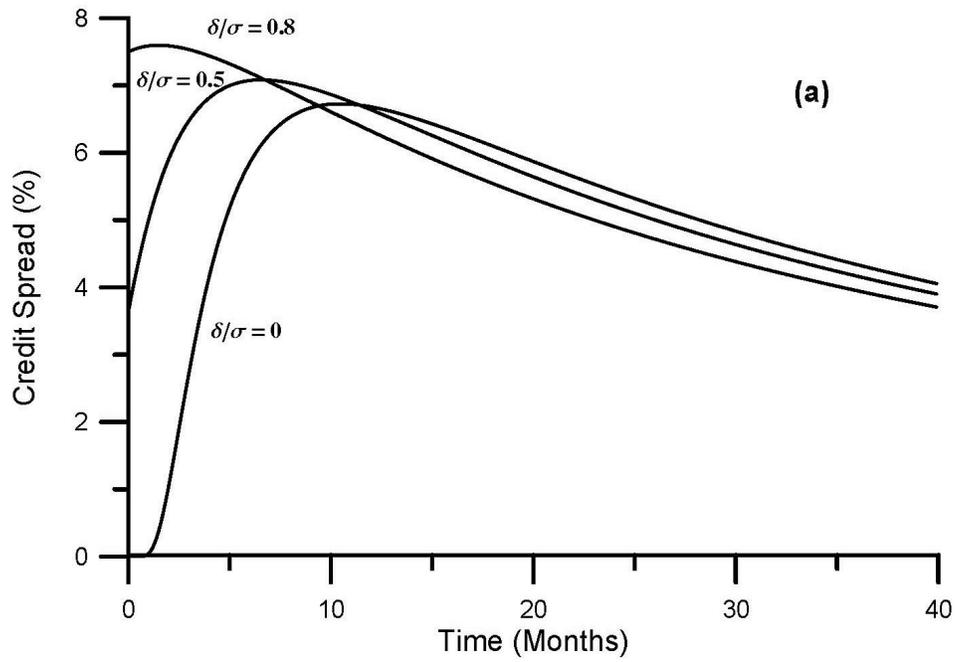

**Figure 3. Calculated short-term structure of credit spreads for bonds of 'B'-rated firms with varying levels of uncertainty in initial distance to default.**

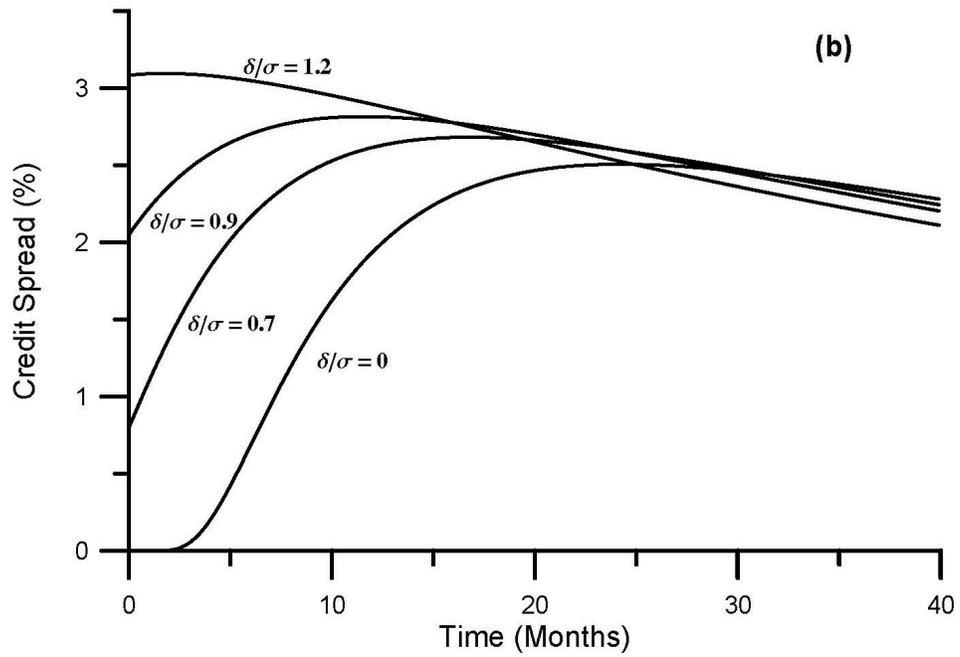

**Figure 3. Calculated short-term structure of credit spreads for bonds of 'BB'-rated firms with varying levels of uncertainty in initial distance to default.**